\documentclass[10pt,twocolumn]{article}
\usepackage[draft]{hyperref}
\usepackage{amsmath}
\usepackage{amssymb}
\usepackage{graphicx}
\usepackage{stfloats}
\usepackage{placeins}
\usepackage{fullpage}
\usepackage{fixltx2e}
\usepackage{soul}
\usepackage[superscript,biblabel]{cite}

    \makeatletter
    \let\@fnsymbol\@arabic %
    \makeatother

\begin{document}
\twocolumn[
  \begin{@twocolumnfalse}

\title{Efficient wide-field FLIM }

\author{Adam J. Bowman\footnotemark[1], Brannon B. Klopfer\footnotemark[1], Thomas Juffmann\footnotemark[2],\:\footnotemark[3]\: and
            Mark A. Kasevich\footnotemark[1]
            }
\maketitle

\noindent\textbf{Nanosecond temporal resolution enables new methods for wide-field imaging like time-of-flight, gated detection, and fluorescence lifetime. The optical efficiency of existing approaches, however, presents challenges for low-light applications common to fluorescence microscopy and single-molecule imaging. We demonstrate the use of Pockels cells for wide-field image gating with nanosecond temporal resolution and high photon collection efficiency. Two temporal frames are obtained by combining a Pockels cell with a pair of polarizing beam-splitters. We show multi-label fluorescence lifetime imaging microscopy (FLIM), single-molecule lifetime spectroscopy, and fast single-frame FLIM at the camera frame rate with $10^3 - 10^5$ times higher throughput than single photon counting. Finally, we demonstrate a space-to-time image multiplexer using a re-imaging optical cavity with a tilted mirror to extend the Pockels cell technique to multiple temporal frames. These methods enable nanosecond imaging with standard optical systems and sensors, opening a new temporal dimension for low-light microscopy.
}
\vspace{3mm}

  \end{@twocolumnfalse}
  
]
\let\thefootnote\relax\footnotetext{\textsuperscript{1}  Physics Department, Stanford University, 382 Via Pueblo Mall, Stanford, California 94305, USA\\ \quad \quad \textsuperscript{2}  Faculty of Physics, University of Vienna, A-1090 Vienna, Austria \\ \quad \quad \textsuperscript{3} Department of Structural and Computational Biology, Max F. Perutz Laboratories, University of Vienna, A-1030 Vienna, Austria. Correspondence and requests for materials should be addressed to A.B. (email: abowman2@stanford.edu)}

\noindent
Existing sensors for wide-field nanosecond imaging sacrifice performance to gain temporal resolution, failing to compete with scientific CMOS and electron-multiplying CCD sensors in low-signal applications. A variety of detectors currently access the nanosecond regime. Gated optical intensifiers (GOIs) based on microchannel plates allow for sub-nanosecond gating in a single image frame, and segmented GOIs can acquire multiple frames when combined with image splitting \cite{Elson2004}. Gating into $n$ frames in this way limits overall collection efficiency to $< 1/n$, and performance is further limited by photocathode quantum efficiency, MCP pixel density, excess noise, and lateral electron drift \cite{Elson2004,Esposito2007,Hirvonen2017,Sparks2017}. Streak camera techniques have also been demonstrated for wide-field imaging, but they also require a photocathode conversion step and additional high-loss encoding \cite{Gao2014,Heshmat}. Silicon photodiode avalanche detector (SPAD) arrays are an emerging solid-state approach, but they are currently limited to sparse fill factors and high dark currents \cite{Burri2014,Ulku2018AFLIM,Hirvonen2017}.

The limitations of current nanosecond imaging techniques are particularly manifest in fluorescence lifetime imaging microscopy (FLIM). Fluorescence lifetime is a sensitive probe of local fluorophore environment and can be used to report factors like pH, polarity, ion concentration, FRET quenching, and viscosity. As lifetime imaging is insensitive to excitation intensity noise, labelling density, and sample photobleaching, it is attractive for many applications. FLIM typically relies on confocal scanning combined with time-correlated single photon counting (TC-SPC) detectors\cite{Berezin2010FluorescenceImaging, Becker2012}.  The throughput of TC-SPC is limited by the detector's maximum count rate (typically 1-100 MHz), and confocal microscopy relies on high excitation intensities that can cause non-linear photodamage to biological samples\cite{Liu2015,Chen2014LatticeResolution}. Frequency domain wide-field approaches are a promising alternative, but they currently require demodulation with either a GOI or high-noise modulated camera chip \cite{Gadella1993FluorescenceScale,Chen2015ModulatedMicroscopy,Raspe}. Given the disadvantages of existing wide-field and TC-SPC approaches, FLIM especially calls for the development of new, efficient imaging strategies to extend its utility for bio-imaging.

Here we demonstrate ultrafast imaging techniques \--- compatible with standard cameras \--- that have no inherent loss or dead time, allowing access to sub-frame rate sample dynamics at timescales as fast as nanosecond fluorescent lifetimes. First, we show an all-photon wide-field imaging system based simply on a pair of polarizing beam-splitters (PBS) and a Pockels cell (PC). This can be used to create two temporal bins or to modulate images on any timescale \--- from nanoseconds to milliseconds. We use this to demonstrate efficient wide-field FLIM of a multi-labelled sample, single molecules, and a biological benchmark. Second, we demonstrate the use of a re-imaging optical cavity as a time-to-space converter to enable $n$-frame ultrafast imaging when combined with a Pockels cell gate. Our approach is photon efficient and retains the sensitivity and image quality of scientific cameras, making it widely compatible and potentially inexpensive. The ability to perform single-frame FLIM without gating loss is a particularly unique advantage, as it enables dynamic FLIM without the loss, noise, and potential motion and intensity artifacts of other approaches.

\begin{figure*}
\centering
  \includegraphics[width=\textwidth]{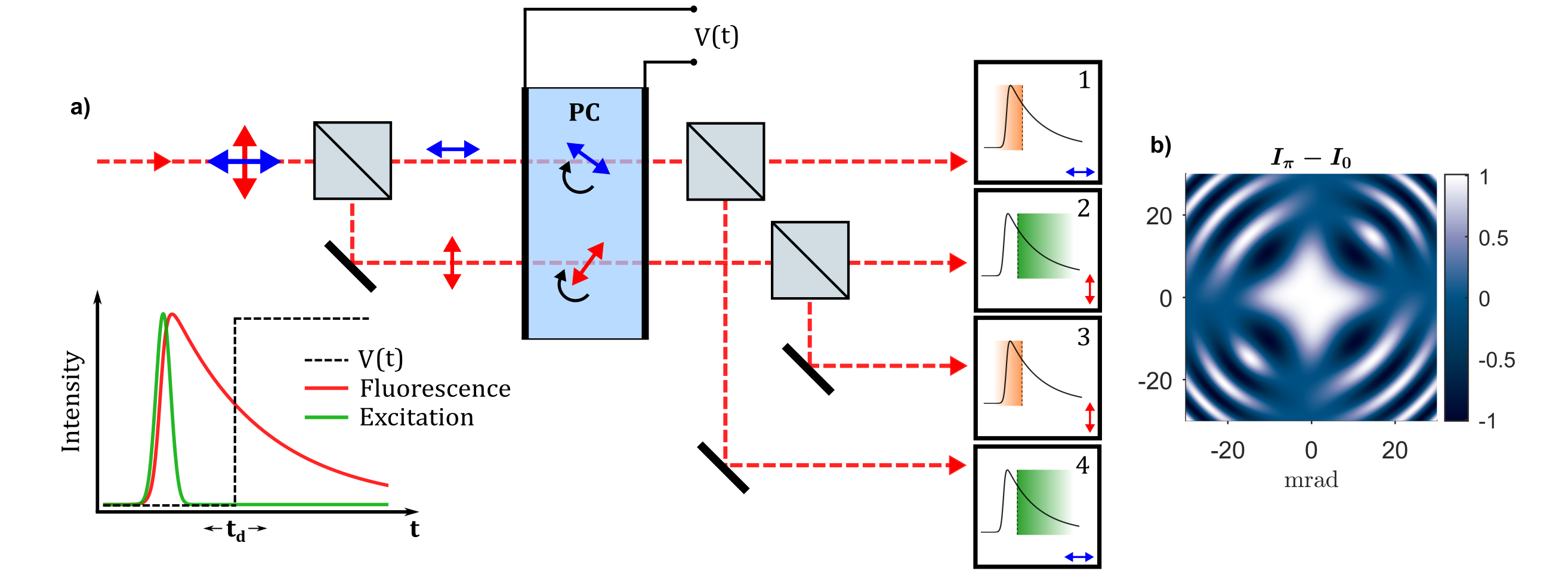}
  \vspace{-8.0mm}
\caption{\label{fig:2bin} {\footnotesize \textbf{Wide-field efficient ultrafast imaging with a Pockels cell }(a) Schematic of two temporal bin wide-field imaging for a single pixel fluorescence decay. Fluorescence emission is first polarized, a time dependent retardance (step function illustrated) is applied by the PC, and polarizations are split again before the sensor. Two pairs of outputs correspond to integrated intensity before (1, 3) and after (2, 4) a step function gate is applied in the illustration. Other modulations V(t) may be applied beyond a simple step function as described in the text. Equal optical path lengths are used in practice. (b) Gating efficiency ($I_{\pi}-I_0$) is calculated for a 30 mm KD*P Pockels cell as a function of incident angle from conscopic interference patterns, demonstrating high efficiency gating for wide-field imaging within 6 mrad half-acceptance angle.
}}
\end{figure*}
\subsection*{Results}
\textbf{Gating with two temporal bins.} Light from an imaging system is polarized with a beam-splitter, and the image associated with each polarization is aligned to propagate through different locations in a wide-aperture PC, as shown in Figure \ref{fig:2bin}. The PC provides an electric field-dependent retardance between the input light's polarization components, mapping the temporal signature of the applied field onto the polarization state of the imaging beams \cite{Davis2014LasersEngineering}. A second PBS after the PC again splits the separated imaging beams, giving four image frames on the camera. The resulting images now encode temporal information, as shown in Figure \ref{fig:2bin}. To illustrate our method, we consider a step function voltage pulse applied at delay time $t_d$ with respect to a short ($\small{\sim}$ ns) excitation pulse. The step function with edge at $t_d$ creates pairs of output images corresponding to integrated signal before and after $t_d$. 

In practice, we implement this configuration with either a Gaussian gating pulse at $t_d$ or a step gate with few nanosecond rise time as described in the following examples. In fact, arbitrary V(t) may be applied to the PC for specific applications (see Discussion). Note that a gating pulse can be applied either as a single shot measurement or over repeated events integrated in one camera frame. Fluorescence lifetime may be recovered by either varying the gate delay $t_d$ to directly measure the fluorescence decay (see multi-label FLIM below) or by single-frame ratios of gated and ungated channel intensities (see single-molecule FLIM below).  In cases where the PC aperture is limited, two separate PC crystals may be used instead of using different areas of the same crystal. Separate gates can be applied to each PC to create four time bins.

\noindent\\
\textbf{Imaging through Pockels cells.} Standard PCs use thick (30-50 mm) potassium dideuterium phosphate (KD*P) crystals with longitudinal field.  These give high extinction ratios and are ubiquitous for Q-switching and phase modulation applications. Off-axis rays experience different birefringent phase shifts than those on-axis, limiting the numerical aperture (NA) of the crystal for wide-field imaging. In an image plane, the PC half angular acceptance $\alpha$ limits the NA of collection optics to $M \alpha$  for small angles, where $M$ is magnification. In a diffraction plane (or infinity corrected space), the field of view (FOV) is instead limited to $2\tan(\alpha) f_\text{obj}$ where $f_{\text{obj}}$ is the imaging objective focal length. For example, we found that a 10 $\mu$m FOV is achieved with a 1.4 NA microscope objective ($f_{\text{obj}}$ = 1.8 mm) and 40 mm thick longitudinal KD*P PC crystal in the infinity space ($\alpha \sim 4$ mrad). FOV can be further improved by magnifying the beam until the PC aperture becomes limiting. Conventional KD*P PCs are limited to long pulse repetition rates in the 10's of kHz by piezoelectric resonances. We note that ultimate repetition rate depends on high voltage pulse shape and crystal dimensions. Electro-optic pulse pickers can operate to 100 kHz and even into MHz rates with low-piezo materials \cite{Kruger1995High-repetition-rateDumping,Kleinbauer200513-WRate, Yan2015100Laser,Bergmann2015MHzCell}. Further, periodic drive avoids exciting piezoelectric resonances and is compatible with frequency-domain FLIM at high excitation rates.

\begin{figure}[t!]
  \includegraphics[width=\columnwidth]{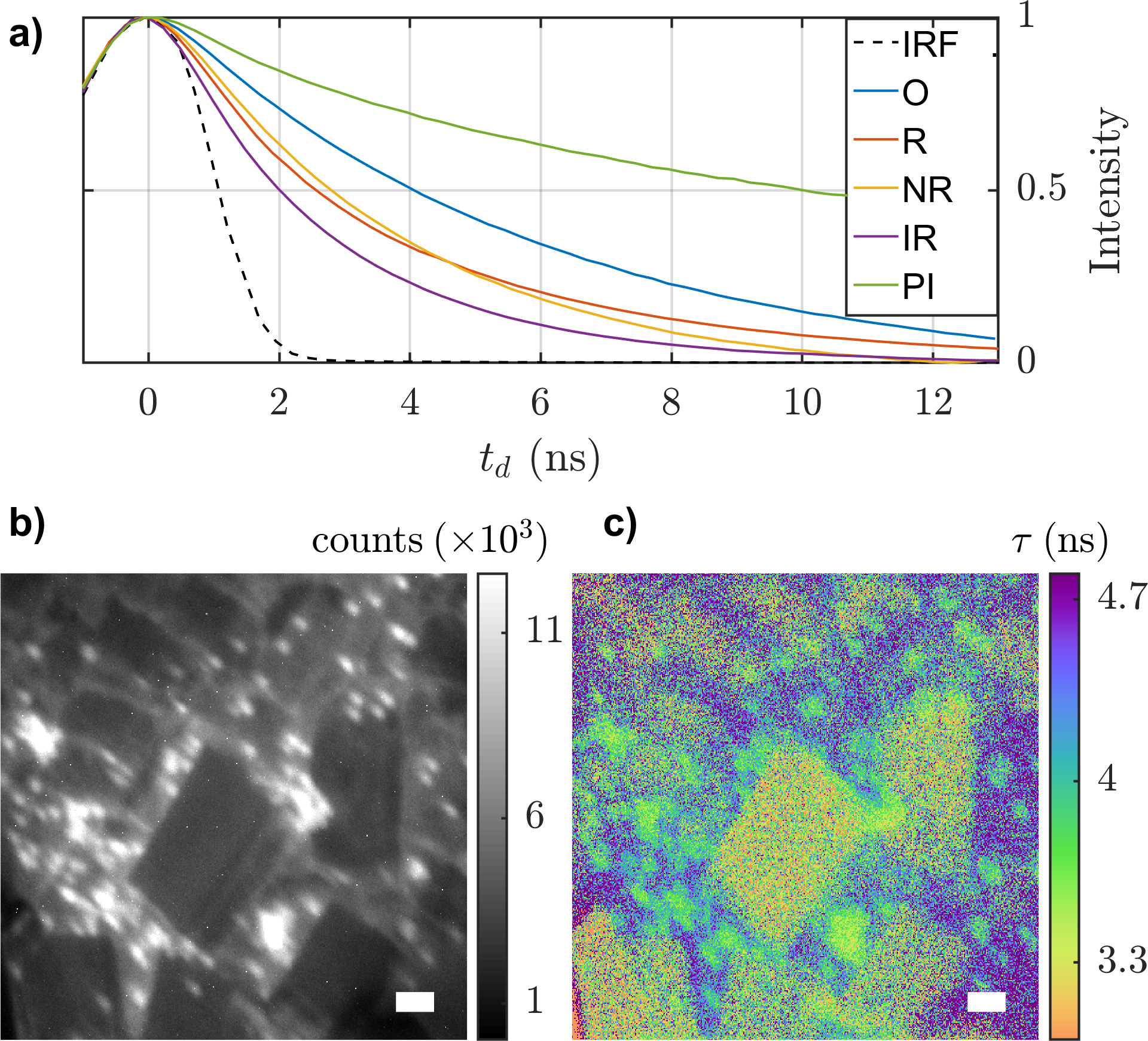}
\caption{\label{fig:traces} {\footnotesize \textbf{Multi-label FLIM }(a) Direct measurement of fluorescence decays obtained by sweeping gate delay time $t_d$ for Orange (O, 4.9 ns), Red (R, 3.4 ns), Nile Red (NR, $\sim$3.1 ns), Infrared (IR, 2.3 ns) (Invitrogen) and Propidium Iodide (PI, 14 ns) (Bangs Laboratories, Inc.) beads. Fitted decay constants $\tau$ are given. The measured Gaussian instrument response function (IRF) is plotted in black.  (b) Intensity image of a three-label wide-field sample of Orange, Nile Red, and Infrared beads (labels strongly overlap spatially) (c) Lifetime image reveals spatial distribution of the labels. Lifetime is measured by fitting the decay traces at each pixel (scale bars 10 $\mu$m).
}}
\end{figure}

\begin{figure*}[ht!]
\centering
  \includegraphics[width=0.9\textwidth]{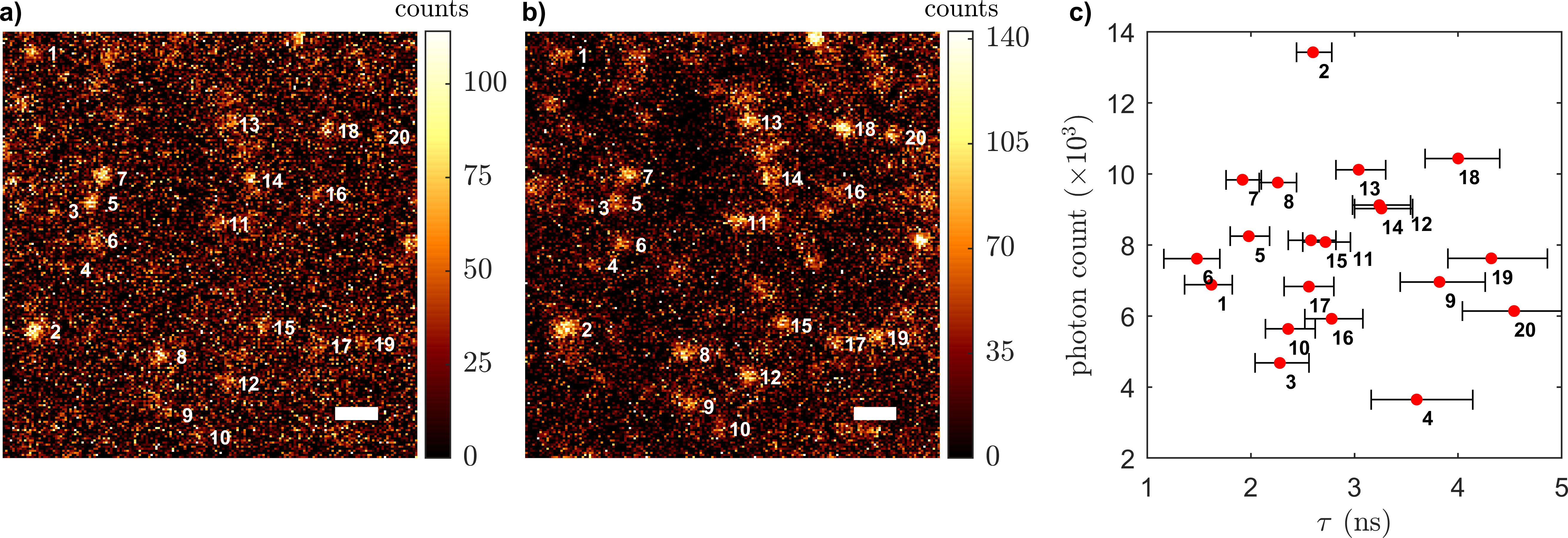}
  \vspace{-5.0mm}
\caption{\label{fig:molecules} {\footnotesize \textbf{Wide-field FLIM of Alexa Fluor 532 molecules} (a) Gated channel intensity (b) Ungated channel intensity (scale bar 1 $\mu$m). (c) Measured lifetimes are plotted along with total brightness for the numbered, diffraction limited regions with SE error bars indicated (see Methods). The majority of these spots are single-molecule emitters as demonstrated by their photobleaching and blinking dynamics (Supplementary Figures 4 and 5). Source data are provided as a Source Data file.
}}
\end{figure*}

\begin{figure*}[ht!]
\centering
  \includegraphics[width=0.9\textwidth]{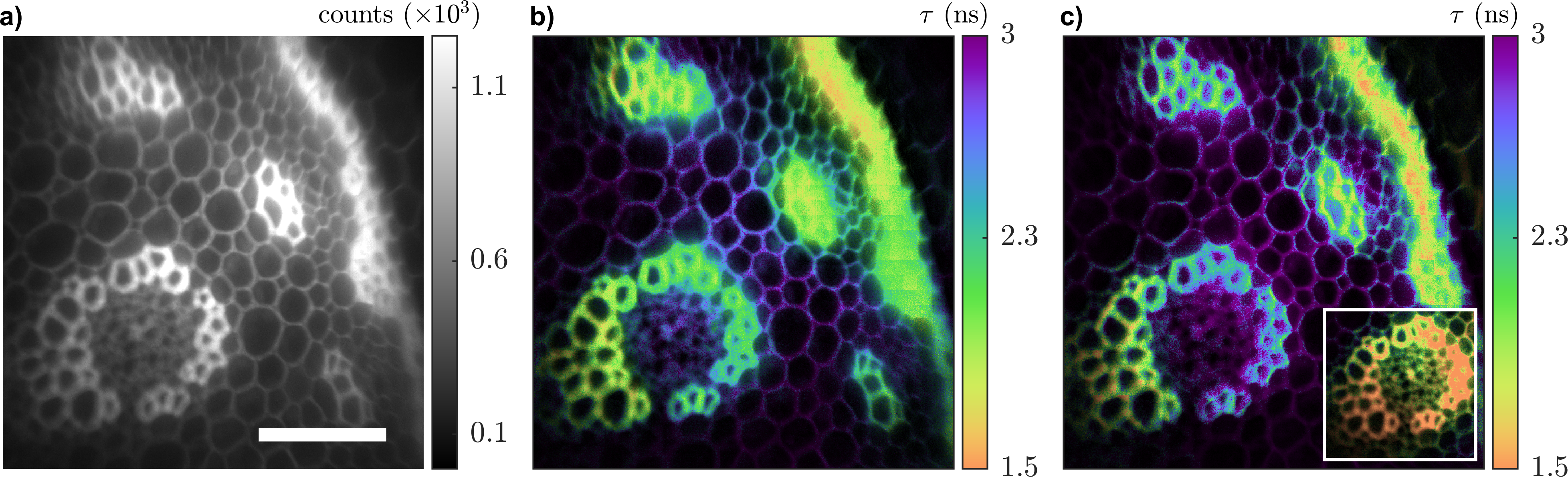}
  \vspace{-0.0mm}
\caption{\label{fig:convallaria} {\footnotesize \textbf{Fast FLIM with a thin PC} (a) Intensity image of \textit{Convallaria majalis} rhizome stained with acridine orange, a standard FLIM benchmark (scale bar 100 $\mu$m). (b) Lifetime image from fitting a timing trace of 100 ms exposures (50 $\mu W$ excitation). (c) Lifetime image from a single 100 ms acquisition frame. Inset demonstrates the same single frame with 2 ms exposure at high excitation intensity (3 mW - high intensity significantly reduced lifetime in this sample, possibly due to photochemistry). Lifetime images include an intensity mask to show sample structure.
}}
\end{figure*}

\begin{figure*} [ht!]
\centering
	\hspace*{-5mm}
  \includegraphics[width=0.8\textwidth]{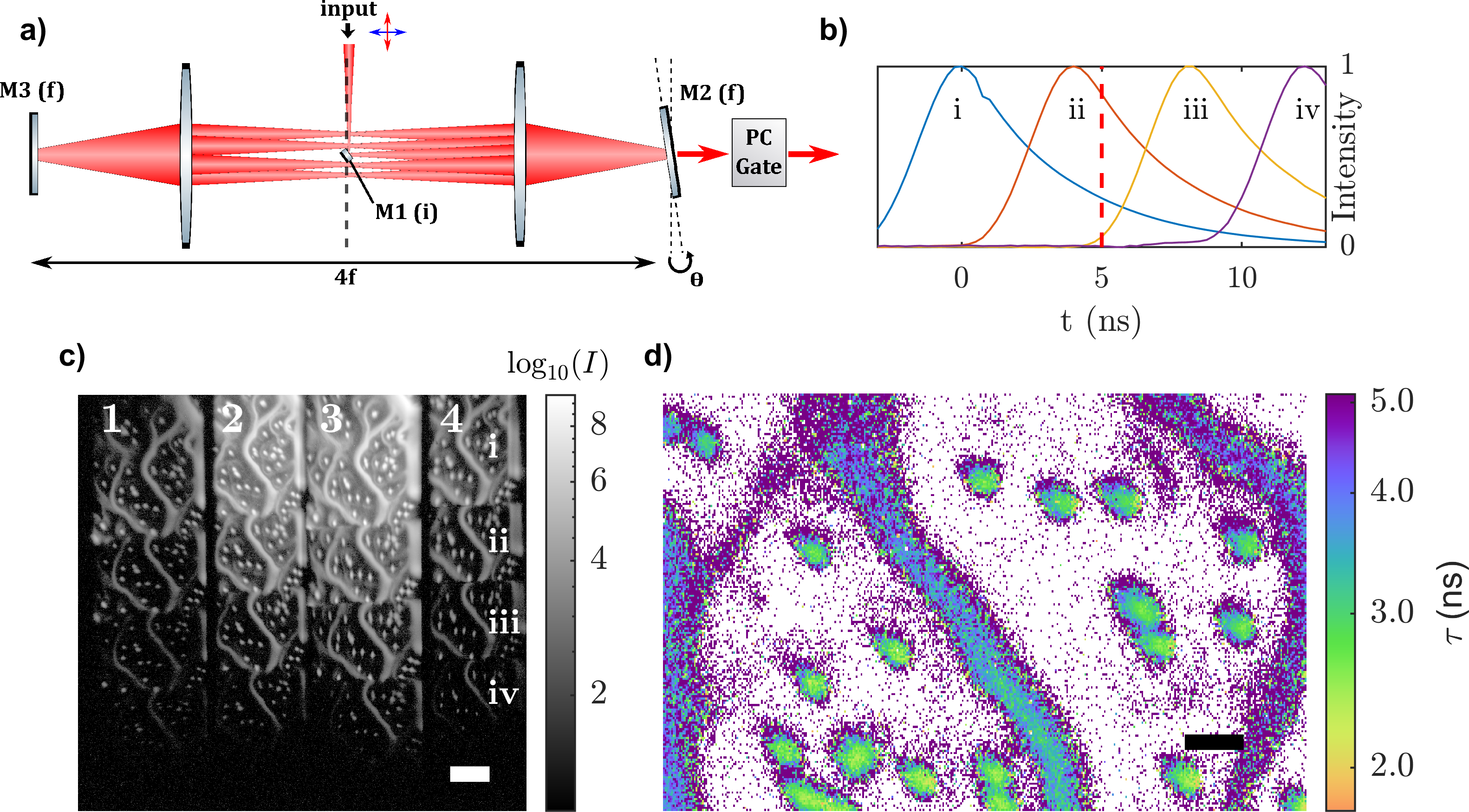}
  
\caption{\label{fig:cavity} {\footnotesize\textbf{ Multi-frame nanosecond imaging with a cavity time-to-space converter }(a) Externally gated, tilted mirror 4$f$ re-imaging cavity. Image input is on small in-coupling mirror M1 in an image plane (i). M2 is tilted at a diffraction plane (f), spatially offsetting the images at the M1 plane each pass. Each round trip, images are passively out-coupled through partially transmissive mirror M2. (b) Normalized image intensities for four output images from the cavity showing a 4 ns round trip delay. (c) Cavity output on camera (CMOS) shows four images output from the PC analyzer for each round trip output from the cavity (columns numbered 1-4 as in Figure \ref{fig:2bin}).  Four round-trips (rows i-iv) are displayed (scale bar 50 $\mu$m). The sample is a mixture of drop-cast Nile Red 2 $\mu$m beads ($\sim$ 3.1 ns) and Orange 0.1 $\mu$m  beads (4.9 ns) that form the diffuse filaments. (d) The ratio of output frames (i,4) and (ii,4) in the gated channel at $t_g = 5$ ns [red line in (b)] is used for single frame FLIM as described in the text.  The two labels are readily differentiated (scale bar 10 $\mu$m). 
}}
\end{figure*}

 To assess gating efficiency, the impact of off-axis birefringence was simulated through the Muller matrix formalism \cite{Bass2010HandbookInstruments} to arrive at a conscopic interference (isogyre) pattern, as viewed through crossed polarizers. Subtracting the transmitted intensity pattern $I$ at zero voltage ($V_0$) from that at the half-wave voltage ($V_{\pi}$) gives the gating efficiency ($I_\pi-I_0$), where the useful NA of the PC is set by the region of high gating efficiency at lower angles [Figure \ref{fig:2bin}(b)]. The PC is treated as a linear homogeneous retarder with off-axis retardance determined by a coordinate transformation of the crystal axes (Supplementary Note 1) \cite{West}. Angular acceptance may be  improved by making the crystal thinner, with a 3 mm crystal increasing $\alpha$ to $\sim$ 20 mrad, effectively removing NA and FOV restrictions. Here we show results using a thick commercial PC (Figures \ref{fig:traces}, \ref{fig:molecules}, and \ref{fig:cavity}) and a custom 3 mm KD*P PC (Figure \ref{fig:convallaria}). Further, complete cancellation of off-axis birefringence may be obtained by combining the negative uniaxial ($n_e < n_o$) KD*P crystal with a positive uniaxial ($n_e > n_o$) static compensating crystal (e.g. MgF$_2$) \cite{West,West2005}. Such a crystal fully compensates for off-axis rays at $V_0$ and further improves the NA at $V_{\pi}$ (KD*P becomes biaxial with applied field, preventing full high voltage compensation). Supplementary Figure 1 compares the effect of off-axis birefringence for thick, thin, and compensated KD*P crystals. \\


\noindent
\textbf{Multi-label FLIM.} The two bin method has no intrinsic gating loss and allows for imaging onto any sensor. Fluorescence lifetime imaging is thus an ideal demonstration for the technique, where the PC gating pulse is applied after delay $t_d$ from the fluorescence excitation. Lifetime may then be determined by either varying the delay time $t_d$ over multiple frames (as used here) or by taking the single-frame ratio of pre- and post-gate intensities (following section). In Figure \ref{fig:traces} we image a mixture of three labels having different lifetimes measured individually to be 3.1 ns (2 $\mu$m Nile Red Invitrogen beads), 4.9 ns (0.1 $\mu$m Orange Invitrogen beads \--- background), and 2.3 ns (0.1 $\mu$m Infrared Invitrogen beads \--- formed into crystals). For this data, the PC was located in the image plane, allowing for wide-field FLIM of bright samples at 0.1 NA and 20x magnification with 100 micron FOV. The sample is excited by laser pulses with duration 1 ns at 532 nm and 5 kHz repetition rate. The fluorescence signal results from the convolution of the decay function with the laser's Gaussian excitation pulse with FWHM pulse width $\small{\sim}2.4\:\sigma_e$. The commercial PC used in Figure \ref{fig:traces} applies a Gaussian gate function $g(t-t_d)$ in our experiment with a pulse width of $2.6$ ns. By sweeping the delay time $t_d$, the convolution of the fluorescence with the gating function is measured: $f(t, \tau,\sigma_e) \ast g(t-t_d)$ .  Temporal information such as fluorescence lifetime may be calculated by directly fitting the measured convolution.  Note that the convolution of excitation and gating functions in this case gives a Gaussian instrument response function (IRF) with $\sigma_\text{IRF} = \sqrt{\sigma_e^2+\sigma_g^2}$, measured directly in Figure \ref{fig:traces}(a). The fitting approach samples the fluorescence decay at more time points and can be advantageous for brightly labeled samples compared to a two-bin measurement . This could be used to more effectively measure multi-exponential decays for instance.\\


\noindent
\textbf{Wide-field FLIM of single molecules.} For signal-limited applications relying on efficient photon collection or requiring fast acquisition rates, fluorescence lifetime is best determined by the ratio of gated and ungated intensity in a single frame.  In Figure \ref{fig:molecules}, we demonstrate wide-field lifetime microscopy of Alexa Fluor 532 (Invitrogen) molecules on glass in a 10 x 10 $\mu$m region. The measured lifetimes are consistent with both the ensemble lifetime of 2.5 ns and the large molecular variation seen in similar studies on glass \cite{Lee2002FluorescenceSurfaces,Xu2017ProbingApproach}. The PC is used in the infinity space of the microscope objective to apply the same Gaussian gating function at $t_d = 1.6 $ ns and 15 kHz repetition rate. The ratio of the gated and ungated intensity is given by
\begin{equation}
R =\frac{\int g(t-t_d)f(t,\tau,\sigma_e) \:dt }{\int f(t,\tau,\sigma_e)[1-g(t-t_d)]\:dt} = \left.\frac{g \ast f}{(1-g)\ast f}\right\rvert_{t=t_d} \:.
\label{ratioint}
\end{equation}
To calculate lifetime, this ratio is experimentally determined by summing intensity in a region of interest around each molecule. This approach allows single-molecule lifetime spectroscopy while maintaining diffraction limited resolution and efficient photon collection  of $\sim 7\times10^3$ photons per molecule (15 s exposure time). Figure \ref{fig:molecules}(c) shows the estimated lifetime and total brightness for each numbered diffraction-limited emitter along with error-bars for the lifetime estimation. Estimation is limited by fluorescence background and dark current here. A low-cost industry CMOS machine vision camera (FLIR) is used for the detector. In this case, the angular acceptance of the PC limits the field of view to 10 $\mu$m but still allows photon collection at 1.4 NA. Single-molecule lifetime spectroscopy in wide field remains challenging with confocal approaches\cite{Luong2005SimultaneousMolecules,Xu2017ProbingApproach,Lee2002FluorescenceSurfaces}, whereas here it is readily demonstrated with PC gating and an inexpensive, high-noise camera.\\\\
\noindent
\textbf{Fast FLIM with a thin PC.}  
By using a thin PC crystal, these techniques are extended to ultra-wide fields of view. A 3 mm thick KD*P Pockels cell with a 20 mm aperture gates nearly the entire output of a standard inverted microscope with an 0.8 NA objective. A $4.5$ ns rising edge pulse (IRF in Supplementary Figure 2) was used at 5 kHz repetition rate to image a standard FLIM benchmark in Figure \ref{fig:convallaria}. Single frame and trace-fitting analysis demonstrate rapid acquisition of megapixel FLIM images with 300 $\mu$m square FOV. Single frame exposures of 100 ms and 2 ms are demonstrated - the latter may be taken at the maximum camera frame rate, enabling future FLIM studies with dynamic samples. These acquisitions show dramatic throughput advantage for wide-field acquisition. The 100 ms and 2 ms frames are acquired at $2300\times$ and $78,000\times$ the speed of 20 MHz TC-SPC in raw photon throughput. The single-frame acquisition method is particularly powerful, as it prevents image motion artifacts (caused by multiple acquisition frames or scanning for example) and allows self-normalization within a single exposure to remove intensity noise. Quantitative lifetimes are easily calculated using the pre-calibrated IRF as described in the prior sections.\\\\
\noindent
\textbf{Gated re-imaging cavities for multi-frame imaging.} Nanosecond imaging with PCs can be extended beyond two temporal bins through the use of gated re-imaging optical cavities. Larger bin numbers enable increased estimation accuracy for multi-exponential decays, improve lifetime dynamic range, and also allow efficient single-shot ultrafast imaging. We exploit the round-trip optical delay of a re-imaging cavity \cite{Arnaud} combined with a tilted cavity mirror to provide nanosecond temporal resolution by spatially separating the cavity round trips. While imaging with $n$-frames using GOIs is limited to $< 1/n$ collection efficiency, this re-imaging cavity technique enables efficient photon collection for low-light or single-photon sensitive applications. In related work, cavities have been used for single-channel orbital angular momentum and wavelength to time conversion\cite{Klopfer2016, Poem2016}. Aligned optical cavities have been used for time-folded optical imaging modalities like multi-pass microscopy  \cite{Juffmann2016a, Heshmat2018PhotographyDimension}.  Our implementation instead employs a re-imaging cavity as the means to obtain temporal resolution for wide-field imaging.

An image is in-coupled to a 4$f$ cavity at the central focal plane by means of a small mirror M1 as shown in Figure \ref{fig:cavity}.  The 4$f$ configuration re-images the end mirrors (diffraction planes) every round trip. If one end mirror M2 is tilted by angle $\theta$, the image position $y_i$ at the central focal plane after $n$  round trips is displaced by $y_i=  f \sin(2 n\theta)$, where $f$ is the focal length of the 4$f$ cavity.  The angle $\theta$ is set such that the resulting images are not blocked by the in-coupling mirror. Each sequentially displaced image is delayed in turn by an additional round trip. To extract temporal information, the spatially separated images need to be either gated externally or simultaneously out-coupled from the cavity using a PC. In the externally gated scheme [schematically shown in Figure \ref{fig:cavity}(a)], light is passively out-coupled each round trip through a transmissive mirror. The spatially displaced images have a relative time delay $\Delta t = 8 f n/c$ based on their number of round trips $n$, and an external gate is simultaneously applied to all delayed images to create temporally distinct frames.  A step function gate V(t) allows lifetime measurement from the ratios of the time-delayed bins, similar to the two-bin case described above. Using the two-bin PC scheme as the external gate gives four image frames from each round trip output [Figure \ref{fig:cavity}(c)]. Photon efficiency, the ratio of detected photons to the number input to the cavity, with end-mirror reflectivity $r$ is given by $1-r^{n}$ after $n$ round trips, ignoring intracavity loss. This efficiency can be made very high for an appropriate choice of $r$. For example, 87\% efficiency is obtained with $r = 0.6$ and $n=4$. It should be noted that the intensity variation between the different frames is caused by partial transmission after $n$ round trips. 

Figure \ref{fig:cavity}(c-d) demonstrates the output from an externally gated tilted mirror cavity. Here a Gaussian gate pulse of width less than the round trip time is used. Lifetime in Figure \ref{fig:cavity}(d) is calculated from the ratio $R$ of two frames [Figure \ref{fig:cavity}(b) images (i,4) and (ii,4)] in the gated channel delayed by one cavity round trip time $t_\text{rt}$ of 4 ns as $R = (g\ast f \rvert_{t_{d}})/ (g\ast f \rvert_{t_{d}+t_\text{rt}})$. Alternatively, both gated and ungated frames could be included in the estimation to make use of all photons as in equation \eqref{ratioint}. It is interesting to compare $n$-bin and two-bin lifetime methods in terms of their theoretical estimation accuracy (see Figure \ref{fig:error}). While the overall accuracies may be closely matched for monoexponentials, $n$-bin methods have the advantage of a wider temporal dynamic range.

 In a second gated cavity scheme (proposed in Supplementary Figure 6), there is instead no transmissive mirror, and all input light is simultaneously out-coupled from the cavity with an intra-cavity Pockels cell and polarizing beamsplitter. Such a scheme directly gives $n$ images with sequential exposures of $ t_\text{rt} = 8f/c $ and leaves no light in the cavity. Either a thin-crystal or compensated PC would be preferable for intra-cavity gating since the light passes through the PC each round trip.

These cavity imaging methods have the advantage of zero dead-time between frames and have no inherent limits on collection efficiency beyond intracavity loss. The externally gated cavity is straightforward to implement with thick-crystal PCs, but has the disadvantage of indirect temporal gating. Intracavity gating instead allows for true $n$-frame ultrafast imaging where each round trip corresponds to one temporally distinct image frame. Round trip times from 1 to 10 ns may be achieved with standard optics. We note that an alternative approach to n-bin imaging could similarly use multiple two-bin gates in series (Figure \ref{fig:2bin}) with the added complexity of multiple PCs and detectors.
\\\\
\noindent
\textbf{Theoretical estimation accuracy.} Two-bin lifetime estimation can perform surprisingly well when compared to the Cram\'er-Rao bound for $n$-bin TC-SPC \cite{Kollnerr1992}. Both two-bin and $n$-bin estimation accuracy scale with photon counting shot noise. Figure \ref{fig:error} shows that $n$-bin measurements have a wider dynamic range of lifetime sensitivity, but that a two-bin PC gate can be nearly as accurate for mono-exponential decays when tuned to the appropriate gate delay. TC-SPC gains a large number of temporal bins from the bit depth of the ADC which dominantly affects the dynamic range. With ideal PC gating, estimation within a factor of two of the shot noise limit (SNL) may be obtained over a decade of lifetimes with peak sensitivity $\sim 1.3 \times$ SNL. In fact, for a step function gate with 1 ns PC rise time, estimation within 2-3$\times$ SNL can be obtained between 1 and 10 ns.

\subsection*{Discussion}
We have presented methods for two and $n$-bin temporal imaging on nanosecond timescales using Pockels cells. Proof-of-concept experiments with single-molecule lifetime spectroscopy and wide-field FLIM demonstrate the potential to bring nanosecond resolution to signal-limited applications. PC imaging methods promise to enhance the acquisition speed and utility of fluorescence lifetime imaging microscopy.

For FLIM applications, nanosecond imaging with PCs enables large improvements in throughput over conventional TC-SPC. Even at low repetition rates, PC FLIM throughput readily surpasses TC-SPC. For example, a PC gated image at a low signal level of 1 photon/pixel/pulse at 15 kHz for a 1 megapixel image would take 7500 times longer to acquire on a 20 MHz confocal TC-SPC system operating at a 10\% count rate (standard to avoid pile-up). This throughput advantage grows linearly with signal and pixel number. Wide-field, high throughput lifetime imaging with PCs could enable imaging of biological dynamics at high frame rate. An example of a relevant application would be real-time imaging of cellular signaling, especially in neurons \cite{Brinks2015Two-PhotonVoltage,Gong2014ImagingSensors,Laviv2016SimultaneousProteins,Raspe}.  FLIM may also be applied as a clinical or in vivo diagnostic and wide-field gating may be readily compatible with endoscopic probes \cite{Munro2005TowardImaging,Sun2010, Sun2013}. We note that frequency modulated cameras have recently been developed to enable high-throughput FLIM, but these suffer from very high dark currents and read noise. PC modulation provides an alternative approach to frequency domain FLIM which can also allow MHz excitation rates.

\begin{figure}
  \includegraphics[width=\columnwidth]{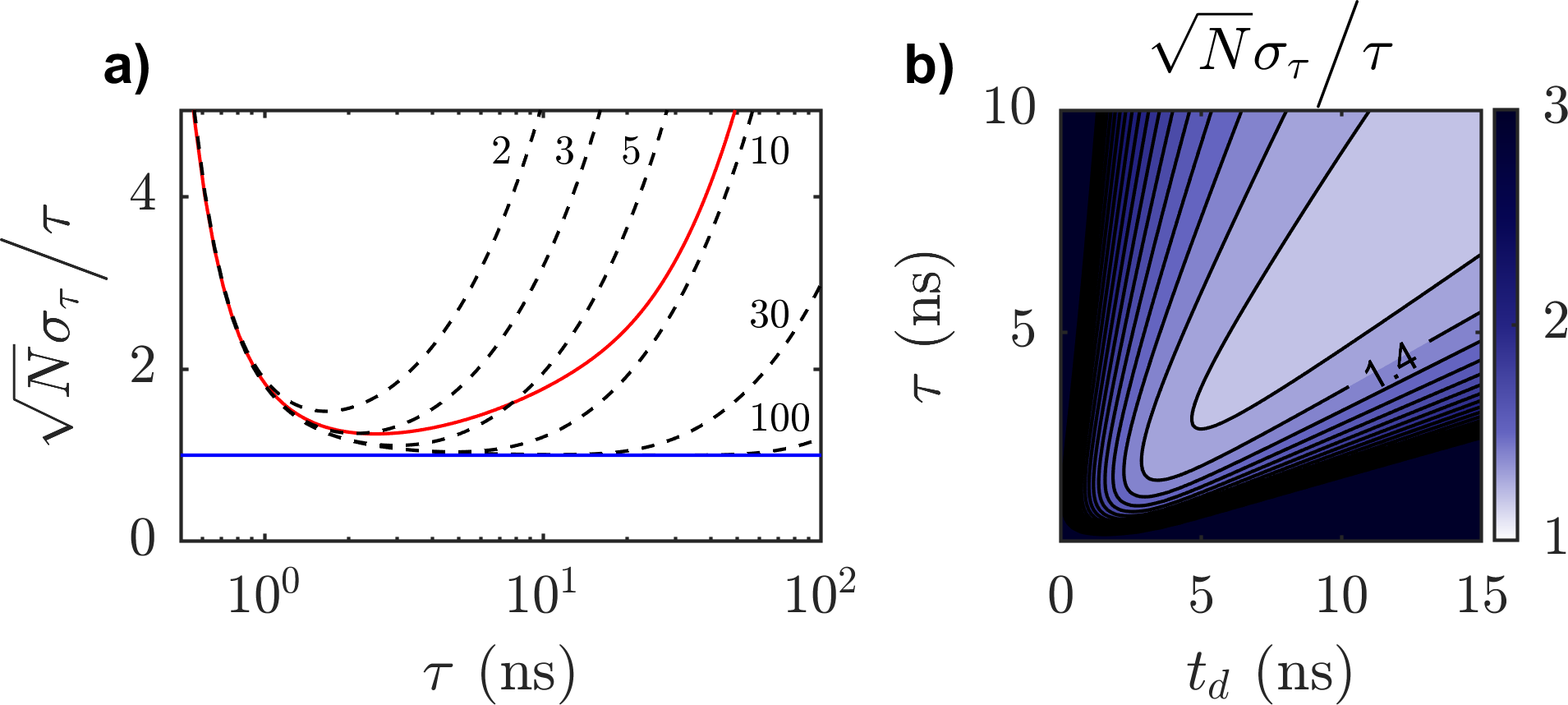}
\caption{\label{fig:error} {\footnotesize\textbf{ Lifetime estimation error} (a) Cram\'er-Rao bound on lifetime estimation accuracy vs. number of bins for a monoexponential fluorescence decay. Dashed lines compare two to $n$-bin lifetime measurements in the case where measurement window is $n$ $\times$ 4 ns cavity round trip time. The red line is a two-bin lifetime simulation at 4 ns $t_d$ without finite measurement window (i.e. ideal step function gate). Note that the range of maximum sensitivity can be shifted with $t_d$. The blue line indicates the shot noise limit: $\sigma_\tau/\tau = 1/\sqrt{N}$. (b) Simulated lifetime resolution for a realistic two-bin PC experiment with a 1 ns 10-90$\%$ logistic rise time PC gate and 1 ns $\sigma_e$, similar to the red line case in (a). Near shot noise limited estimation accuracy is obtained for $\tau >$ PC rise time. 
}}
\end{figure}

PC gating may further allow for new microscopy techniques by exploiting the nanosecond temporal dimension. For example, spectral information has been used to enable multi-labeling of biological samples, which proves important in understanding complex intracellular interactions\cite{Valm2017}. Fluorescence lifetime may similarly provide an attractive temporal approach for unmixing multi-labeled signals \cite{Fan2018Lifetime-engineeredImaging}. Confocal FLIM has already been applied to this problem\cite{Niehorster2016a}.  In studying single molecules, the capability to combine parallel lifetime measurements with spatial and spectral channels could allow for new types of high-throughput spectroscopy experiments to study molecular populations and photophysical states \cite{Wang2013LifetimeTrap,Zhang2015Ultrahigh-throughputMicroscopy,Mlodzianoski2016Super-ResolutionFluctuations,Squires2017DirectC-phycocyanin,Yan2018SpectrallySpectroscopy}. New information from lifetime could also be used to enhance spatial localization in super-resolution microscopy \cite{Dong2016}. Further, temporal gating could be used to suppress background autofluorescence occurring at short lifetimes \cite{Cordina2018ReducedChelates}.

While we have primarily focused on applications in fluorescence microscopy, we also note that PC nanosecond imaging techniques could be more broadly applied in quantum optics for fast gating, lock-in detection, event selection, or multi-pass microscopy \cite{Fickler2016Quantum10010,Klopfer2016,Juffmann2016a}. Other useful operation modes may be realized with the two-bin PC scheme by applying different modulations V(t). For example, a linear ramp of V(t) creates a unique mapping of time to output intensity to temporally localize photon bursts (e.g. molecule blinking) by polarization streaking (see Supplementary Figure 3). Periodic V(t) could also be used to implement wide-field lock-in detection. Traditional fast-imaging applications in plasma physics, LIBS spectroscopy, combustion, time-of-flight techniques, and fluid dynamics could also benefit from sensitive single-shot imaging \cite{Hahn2012Laser-InducedFields,Gao2016,Liang2018Single-shotImaging,Heshmat2018PhotographyDimension,Heshmat}. The $n$-frame tilted mirror re-imaging cavity is particularly unique in its ability to perform single-shot ultrafast imaging of weak, non-repetitive events with zero deadtime between frames when using an internal PC gate (Supplementary Figure 6). It could also prove useful for wide dynamic range lifetime imaging.  We note that strategies for time-of-flight imaging and LIDAR have also been recently demonstrated using PC modulation for the timing of reflected light pulses \cite{Jo, Zhang2017HighModulation, Chen2018Polarization-modulatedModulator}. 

In summary, wide-field PC FLIM was demonstrated in single-frame and time trace modalities. Single-molecule lifetime spectroscopy showed compatibility with signal limited applications. By using a thin PC crystal, the technique was extended to ultra wide-field FLIM with single frame acquisition. FLIM images were acquired on a standard biological benchmark with exposure times down to 2 ms and acquisition speeds to the camera frame rate. Finally, a new method using re-imaging cavities to enable ultrafast imaging by time-to-space multiplexing was shown. These techniques promise to open the nanosecond regime to signal-limited applications like wide-field and single-molecule fluorescence microscopy. Further, they are broadly compatible with any imaging system and sensor, giving potential applications in a variety of fields.

\bigskip
\bigskip
\bigskip
\bigskip
\noindent\textbf{\large Methods}

\noindent\footnotesize{\textbf{Experimental setup.} 
FLIM was performed with a homemade fluorescence microscope for figures \ref{fig:traces}, \ref{fig:molecules}, and \ref{fig:cavity}. A Nikon PlanApo 100x VC 1.4 NA oil-immersion objective was used for single-molecule microscopy. All other data was taken with a 20x, 0.8 NA Zeiss PlanApo objective.  Excitation pulses (1 ns FWHM) at 532 nm were generated by a Q-switched Nd:YAG at 5 kHz repetition (15 kHz for single-molecule data) (Standa Q10-SH).  The detector was a machine vision CMOS camera (FLIR BFS-U3-32S4M-C). A 10 mm aperture, 40 mm thick dual crystal longitudinal KD*P PC embedded in a 50 $\Omega$ transmission line was used (Lasermetrics 1072). High voltage gating pulses were generated into 50 $\Omega$ with an amplitude of 1.3 kV, 2.8 ns FWHM  (FID GmbH) attaining 85$\%$ of $V_{\pi}$ and $\sigma_{\text{IRF}} = 1.1$ ns. Laser and HV pulser were synchronized with a DG 535 delay generator (Stanford Research Systems). Timing jitter was $< 100$ ps. Long transmission lines were used to prevent spurious pulse reflections during fluorescence decay.  For single-molecule data, only two of the output frames (one output from first PBS) were used to maximize FOV through the PC, limiting photon efficiency to $\sim 50\%$. This is not a fundamental limitation of the technique but was used to simplify our implementation with a limited single PC aperture. IRFs were acquired using a frosted glass sample.\\\\
\noindent
The thin crystal demonstration was performed on an inverted microscope (Zeiss Axiovert) using a 20x, 0.8 NA Zeiss PlanApo objective and Andor Neo5.5 sCMOS. A 3 mm thick, 20 mm aperture KD*P longitudinal PC was home-built along with a high voltage driver capable of supplying nanosecond switching pulses with amplitudes up to 5 kV. A gating efficiency of 0.8 is used with a rise time of 4.5 ns for the data in Figure \ref{fig:convallaria}. Only one polarization channel is demonstrated here. Both channels may be incorporated by adding a second PC or Wollaston prism.\\\\
\noindent
The 4$f$ re-imaging cavity used for the $n$-bin demonstration used a 3 mm prism mirror (Thorlabs MRA03-G01) for in-coupling and $f$ = 150 mm ($t_\text{rt} = 8f/c = 4.0$ ns). Passive out-coupling was through a neutral density filter of optical density 1 ($R =  0.4$ and $T = 0.1$). Relay lenses were used to create an image plane at the PC and again at the camera (CMOS). Pick-off mirrors combined imaging beams generated by the two PBS with equal path lengths.\\\\
}
\noindent\footnotesize{\textbf{Sample preparation.} 
Alexa 532 single-molecule samples were prepared by drop casting dilute solution onto a hydrophobic substrate, then placing and removing a pristine coverslip. A dense field was photobleached to the point that single, diffraction-limited emitters were observed. Step-like photobleaching is shown in the Supplementary Figures 4 and 5 along with blink-on dynamics. While multi-molecule emission within a diffraction limited spot was certainly also seen, a majority of the emitters were single molecules. Fluorescent bead samples were prepared by drop-casting directly onto coverglass. Invitrogen IR bead solution formed crystals as seen in Figure \ref{fig:traces}.
}

\bigskip
\noindent\footnotesize{\textbf{Data analysis.}
Lifetimes were computed by both ratiometric calculation from image intensities and by time-trace fitting. In ratiometric calculation, a numerically generated lookup table is used to convert between the measured ratio and estimated lifetime according to the equations in the text and the pre-characterized IRF. Due to our specific $t_d$ and Gaussian gate pulse in Figure \ref{fig:molecules}, lifetimes below 1.1 ns are redundant with those above 1.1 ns in the numerical conversion. We report the larger lifetime value. In timing trace calculation, fitting by least squares was used to estimate lifetime.  The PC applies a time-varying retardance to linearly polarized input as $\delta = 2 \pi r_{63} V n_o^3/\lambda$, where the birefringent phase shift $\delta$ is determined by the applied voltage V, ordinary index of refraction $n_o$, and the longitudinal electro-optic coefficient $r_{63}$. Transmission in the parallel and perpendicular beamsplitter channels is $T_\parallel = \sin^2(\delta/2)$ and $T_\perp = \cos^2(\delta/2)$. Lifetime calculations account for the imperfect gating efficiency of the Pockels cell as captured in the IRF. In figures \ref{fig:traces}, \ref{fig:molecules}, and \ref{fig:cavity}, a constant IRF is assumed across a conservative FOV. This may cause position-dependent lifetime errors. In Figure \ref{fig:convallaria}, spatial variation is more apparent due to large FOV and is included in lifetime calculations. A beamsplitter in the microscope filter slider allows rapid switching between fluorescence and frosted glass IRF calibration. IRF calibration may also be performed with a short lifetime dye.  \\\\
\noindent
Single-molecule gated and ungated intensities were determined by summing $N_p$ pixels corresponding to each molecule region of interest after background subtraction. Error bars in Figure \ref{fig:molecules}(c) account for shot noise in the gated (G) and ungated (U) frames and for the background standard deviations $\sigma_g$ and $\sigma_{u}$ in the ratio SE $\sigma_R$ as
\begin{equation*}
\sigma_R = \frac{G}{U}\sqrt{\frac{1}{G} + \frac{1}{U }+ N_p(\frac{\sigma_g^2}{G^2}+\frac{\sigma_{u}^2}{U^2})}\:.
\end{equation*}
\noindent
Background is the dominant error term here combining background signal with a high camera dark current.\\\\
\noindent
The Cram\'er-Rao bound for $n$-bin lifetime estimation in a fixed time window of width $T$ may be directly calculated from a multinomial probability distribution \cite{Kollnerr1992}. Fixed window bounds in Figure \ref{fig:error} were found by setting $T = n \times t_{rt} $ for $n$ round trips. The photon normalized Cram\'er-Rao bound for $n$ bins is
\begin{equation*}
\frac{\sqrt N\sigma_\tau}{\tau} = \frac{n \tau}{T}\sqrt{(1-e^{-T/\tau})}\bigg[
\frac{e^{{\frac{T}{n\tau}}}(1-e^{-T/\tau})}
{(e^{\frac{T}{n\tau}}-1)^2} -
\frac{n^2}
{e^{T/\tau}-1}\bigg]^{-1/2}\:.
\end{equation*}
}

\bigskip
\noindent\textbf{\large Acknowledgements}
We thank Yonatan Israel and Rachel Gruenke. This research was funded by the Gordon and Betty Moore Foundation. AB and BK acknowledge support from the Stanford Graduate Fellowship. AB acknowledges support from the National Science Foundation Graduate Research Fellowship Program under grant 1656518.
\noindent\footnotesize{}

\bigskip
\noindent\textbf{\large Author contributions}
AB conceived the idea and performed the experiments. BK and TJ performed initial work on PC gating. AB and MK prepared the manuscript.
\noindent 

\bigskip
\noindent\textbf{\large Data availability}

\noindent{The data that support the findings of this study are available from the corresponding author upon request.}

\bigskip
\noindent\textbf{\large Competing financial interests}

\noindent{The authors declare no competing financial interests.}


\bibliographystyle{naturemag}
\bibliography{mendeley}

\begin{thebibliography}{10}
\expandafter\ifx\csname url\endcsname\relax
  \def\url#1{\texttt{#1}}\fi
\expandafter\ifx\csname urlprefix\endcsname\relax\def\urlprefix{URL }\fi
\providecommand{\bibinfo}[2]{#2}
\providecommand{\eprint}[2][]{\url{#2}}

\bibitem{Elson2004}
\bibinfo{author}{Elson, D.~S.} \emph{et~al.}
\newblock \bibinfo{title}{{Real-time time-domain fluorescence lifetime imaging
  including single-shot acquisition with a segmented optical image
  intensifier}}.
\newblock \emph{\bibinfo{journal}{New Journal of Physics}}
  \textbf{\bibinfo{volume}{6}} (\bibinfo{year}{2004}).

\bibitem{Esposito2007}
\bibinfo{author}{Esposito, A.}, \bibinfo{author}{Gerritsen, H.~C.} \&
  \bibinfo{author}{Wouters, F.~S.}
\newblock \bibinfo{title}{{Optimizing frequency-domain fluorescence lifetime
  sensing for high-throughput applications: photon economy and acquisition
  speed}}.
\newblock \emph{\bibinfo{journal}{J. Opt. Soc. Am. A}}
  \textbf{\bibinfo{volume}{24}}, \bibinfo{pages}{3261--3273}
  (\bibinfo{year}{2007}).

\bibitem{Hirvonen2017}
\bibinfo{author}{Hirvonen, L.~M.} \& \bibinfo{author}{Suhling, K.}
\newblock \bibinfo{title}{{Wide-field TCSPC: methods and applications}}.
\newblock \emph{\bibinfo{journal}{Meas. Sci. Technol}}
  \textbf{\bibinfo{volume}{28}}, \bibinfo{pages}{012003}
  (\bibinfo{year}{2017}).

\bibitem{Sparks2017}
\bibinfo{author}{Sparks, H.} \emph{et~al.}
\newblock \bibinfo{title}{{Characterisation of new gated optical image
  intensifiers for fluorescence lifetime imaging}}.
\newblock \emph{\bibinfo{journal}{Review of Scientific Instruments}}
  \textbf{\bibinfo{volume}{88}}, \bibinfo{pages}{013707}
  (\bibinfo{year}{2017}).

\bibitem{Gao2014}
\bibinfo{author}{Gao, L.}, \bibinfo{author}{Liang, J.}, \bibinfo{author}{Li,
  C.} \& \bibinfo{author}{Wang, L.~V.}
\newblock \bibinfo{title}{{Single-shot compressed ultrafast photography at one
  hundred billion frames per second}}.
\newblock \emph{\bibinfo{journal}{Nature}} \textbf{\bibinfo{volume}{516}},
  \bibinfo{pages}{74--77} (\bibinfo{year}{2014}).

\bibitem{Heshmat}
\bibinfo{author}{Heshmat, B.}, \bibinfo{author}{Satat, G.},
  \bibinfo{author}{Barsi, C.} \& \bibinfo{author}{Raskar, R.}
\newblock \bibinfo{title}{{Single-shot ultrafast imaging using parallax-free
  alignment with a tilted lenslet array}}.
\newblock In \emph{\bibinfo{booktitle}{CLEO: Science and Innovations}},
  \bibinfo{pages}{pp. STu3E--7} (\bibinfo{year}{2014}).

\bibitem{Burri2014}
\bibinfo{author}{Burri, S.} \emph{et~al.}
\newblock \bibinfo{title}{{A 65k pixel, 150k frames-per-second camera with
  global gating and micro-lenses suitable for fluorescence lifetime imaging}}.
\newblock In \emph{\bibinfo{booktitle}{Proc. SPIE 9141, Optical Sensing and
  Detection III}}, \bibinfo{pages}{914109} (\bibinfo{year}{2014}).

\bibitem{Ulku2018AFLIM}
\bibinfo{author}{Ulku, A.~C.} \emph{et~al.}
\newblock \bibinfo{title}{{A 512 x 512 SPAD Image Sensor with Integrated Gating
  for Widefield FLIM}}.
\newblock \emph{\bibinfo{journal}{IEEE Journal of Selected Topics in Quantum
  Electronics}} \textbf{\bibinfo{volume}{25}}, \bibinfo{pages}{1--12}
  (\bibinfo{year}{2018}).

\bibitem{Berezin2010FluorescenceImaging}
\bibinfo{author}{Berezin, M.~Y.} \& \bibinfo{author}{Achilefu, S.}
\newblock \bibinfo{title}{{Fluorescence Lifetime Measurements and Biological
  Imaging}}.
\newblock \emph{\bibinfo{journal}{Chemical Reviews}}
  \textbf{\bibinfo{volume}{110}}, \bibinfo{pages}{2641--2684}
  (\bibinfo{year}{2010}).

\bibitem{Becker2012}
\bibinfo{author}{Becker, W.}
\newblock \bibinfo{title}{{Fluorescence lifetime imaging - techniques and
  applications}}.
\newblock \emph{\bibinfo{journal}{Journal of Microscopy}}
  \textbf{\bibinfo{volume}{247}}, \bibinfo{pages}{119--136}
  (\bibinfo{year}{2012}).

\bibitem{Liu2015}
\bibinfo{author}{Liu, Z.}, \bibinfo{author}{Lavis, L.~D.} \&
  \bibinfo{author}{Betzig, E.}
\newblock \bibinfo{title}{{Imaging Live-Cell Dynamics and Structure at the
  Single-Molecule Level}}.
\newblock \emph{\bibinfo{journal}{Molecular Cell}}
  \textbf{\bibinfo{volume}{58}}, \bibinfo{pages}{644--659}
  (\bibinfo{year}{2015}).

\bibitem{Chen2014LatticeResolution}
\bibinfo{author}{Chen, B.-C.} \emph{et~al.}
\newblock \bibinfo{title}{{Lattice light-sheet microscopy: Imaging molecules to
  embryos at high spatiotemporal resolution}}.
\newblock \emph{\bibinfo{journal}{Science}} \textbf{\bibinfo{volume}{346}},
  \bibinfo{pages}{1257998} (\bibinfo{year}{2014}).

\bibitem{Gadella1993FluorescenceScale}
\bibinfo{author}{Gadella, T. W.~J.}, \bibinfo{author}{Jovin, T.~M.} \&
  \bibinfo{author}{Clegg, R.~M.}
\newblock \bibinfo{title}{{Fluorescence lifetime imaging microscopy (FLIM)
  Spatial resolution of microstructures on the nanosecond time scale}}.
\newblock \emph{\bibinfo{journal}{Biophysical Chemistry}}
  \textbf{\bibinfo{volume}{48}}, \bibinfo{pages}{221--239}
  (\bibinfo{year}{1993}).

\bibitem{Chen2015ModulatedMicroscopy}
\bibinfo{author}{Chen, H.}, \bibinfo{author}{Holst, G.} \&
  \bibinfo{author}{Gratton, E.}
\newblock \bibinfo{title}{{Modulated CMOS Camera for Fluorescence Lifetime
  Microscopy}}.
\newblock \emph{\bibinfo{journal}{Microsc. Res. Tech}}
  \textbf{\bibinfo{volume}{78}}, \bibinfo{pages}{1075--1081}
  (\bibinfo{year}{2015}).

\bibitem{Raspe}
\bibinfo{author}{Raspe, M.} \emph{et~al.}
\newblock \bibinfo{title}{{siFlim: single-image frequency-domain FLIM provides
  fast and photon- efficient lifetime data}}.
\newblock \emph{\bibinfo{journal}{Nature Methods}}
  \textbf{\bibinfo{volume}{13}}, \bibinfo{pages}{501--504}
  (\bibinfo{year}{2016}).

\bibitem{Davis2014LasersEngineering}
\bibinfo{author}{Davis, C.~C.}
\newblock \emph{\bibinfo{title}{{Lasers and electro-optics: fundamentals and
  engineering}}} (\bibinfo{publisher}{Cambridge University Press},
  \bibinfo{year}{2014}).

\bibitem{Kruger1995High-repetition-rateDumping}
\bibinfo{author}{Kr{\"{u}}ger, E.} \& \bibinfo{author}{Krijger, E.}
\newblock \bibinfo{title}{{High-repetition-rate electro-optic cavity dumping}}.
\newblock \emph{\bibinfo{journal}{Review of Scientific Instruments}}
  \textbf{\bibinfo{volume}{66}}, \bibinfo{pages}{1028} (\bibinfo{year}{1995}).

\bibitem{Kleinbauer200513-WRate}
\bibinfo{author}{Kleinbauer, J.}, \bibinfo{author}{Knappe, â.~R.} \&
  \bibinfo{author}{Wallenstein, R.}
\newblock \bibinfo{title}{{13-W picosecond Nd:GdVO 4 regenerative amplifier
  with 200-kHz repetition rate}}.
\newblock \emph{\bibinfo{journal}{Appl. Phys. B}}
  \textbf{\bibinfo{volume}{81}}, \bibinfo{pages}{163--166}
  (\bibinfo{year}{2005}).

\bibitem{Yan2015100Laser}
\bibinfo{author}{Yan, R.} \emph{et~al.}
\newblock \bibinfo{title}{{100 kHz, 3.1 ns, 1.89 J cavity-dumped burst-mode
  Nd:YAG MOPA laser}}.
\newblock \emph{\bibinfo{journal}{Optics Express}}
  \textbf{\bibinfo{volume}{5}}, \bibinfo{pages}{761--769}
  (\bibinfo{year}{2015}).

\bibitem{Bergmann2015MHzCell}
\bibinfo{author}{Bergmann, F.} \emph{et~al.}
\newblock \bibinfo{title}{{MHz Repetion Rate Yb:YAG and Yb:CaF2 Regenerative
  Picosecond Laser Amplifiers with a BBO Pockels Cell}}.
\newblock \emph{\bibinfo{journal}{Applied Sciences}}
  \textbf{\bibinfo{volume}{5}}, \bibinfo{pages}{761--769}
  (\bibinfo{year}{2015}).

\bibitem{Bass2010HandbookInstruments}
\bibinfo{author}{Bass, M.}
\newblock \emph{\bibinfo{title}{{Handbook of Optics, Volume 1: Geometric and
  Physical Optics, Polarized Light, Components and Instruments}}}
  (\bibinfo{publisher}{McGraw-Hill}, \bibinfo{year}{2010}).

\bibitem{West}
\bibinfo{author}{West, E.~A.}
\newblock \bibinfo{title}{{Extending the field of view of KD*P electrooptic
  modulators}}.
\newblock \emph{\bibinfo{journal}{Applied Optics}}
  \textbf{\bibinfo{volume}{17}}, \bibinfo{pages}{3010--3013}
  (\bibinfo{year}{1978}).

\bibitem{West2005}
\bibinfo{author}{West, E.~A.}, \bibinfo{author}{Gary, G.~A.},
  \bibinfo{author}{Noble, M.}, \bibinfo{author}{Choudhary, D.} \&
  \bibinfo{author}{Robinson, B.}
\newblock \bibinfo{title}{{Large field-of-view KD*P modulator for solar
  polarization measurements}}.
\newblock In \emph{\bibinfo{booktitle}{Proc. SPIE 5888, Polarization Science
  and Remote Sensing II}}, \bibinfo{pages}{588806} (\bibinfo{year}{2005}).

\bibitem{Lee2002FluorescenceSurfaces}
\bibinfo{author}{Lee, M.}, \bibinfo{author}{Kim, J.}, \bibinfo{author}{Tang,
  J.} \& \bibinfo{author}{Hochstrasser, R.~M.}
\newblock \bibinfo{title}{{Fluorescence quenching and lifetime distributions of
  single molecules on glass surfaces}}.
\newblock \emph{\bibinfo{journal}{Chemical Physics Letters}}
  \textbf{\bibinfo{volume}{359}}, \bibinfo{pages}{412--419}
  (\bibinfo{year}{2002}).

\bibitem{Xu2017ProbingApproach}
\bibinfo{author}{Xu, B.} \emph{et~al.}
\newblock \bibinfo{title}{{Probing the inhomogeneity and intermediates in the
  photosensitized degradation of rhodamine B by Ag3PO4 nanoparticles from an
  ensemble to a single molecule approach}}.
\newblock \emph{\bibinfo{journal}{RSC Adv.}} \textbf{\bibinfo{volume}{7}},
  \bibinfo{pages}{40896} (\bibinfo{year}{2017}).

\bibitem{Luong2005SimultaneousMolecules}
\bibinfo{author}{Luong, A.~K.}, \bibinfo{author}{Gradinaru, C.~C.},
  \bibinfo{author}{Chandler, D.~W.} \& \bibinfo{author}{Hayden, C.~C.}
\newblock \bibinfo{title}{{Simultaneous Time-and Wavelength-Resolved
  Fluorescence Microscopy of Single Molecules}}.
\newblock \emph{\bibinfo{journal}{J. Phys. Chem. B.}}
  \textbf{\bibinfo{volume}{109}}, \bibinfo{pages}{15691--15698}
  (\bibinfo{year}{2005}).

\bibitem{Arnaud}
\bibinfo{author}{Arnaud, J.~A.}
\newblock \bibinfo{title}{{Degenerate Optical Cavities}}.
\newblock \emph{\bibinfo{journal}{Applied Optics}}
  \textbf{\bibinfo{volume}{8}}, \bibinfo{pages}{189--196}
  (\bibinfo{year}{1969}).

\bibitem{Klopfer2016}
\bibinfo{author}{Klopfer, B.~B.}, \bibinfo{author}{Juffmann, T.} \&
  \bibinfo{author}{Kasevich, M.~A.}
\newblock \bibinfo{title}{{Iterative creation and sensing of twisted light}}.
\newblock \emph{\bibinfo{journal}{Optics Letters}}
  \textbf{\bibinfo{volume}{41}}, \bibinfo{pages}{5744--5747}
  (\bibinfo{year}{2016}).

\bibitem{Poem2016}
\bibinfo{author}{Poem, E.}, \bibinfo{author}{Hiemstra, T.},
  \bibinfo{author}{Eckstein, A.}, \bibinfo{author}{Jin, X.-M.} \&
  \bibinfo{author}{Walmsley, I.~A.}
\newblock \bibinfo{title}{{Free-space spectro-temporal and spatio-temporal
  conversion for pulsed light}}.
\newblock \emph{\bibinfo{journal}{Optics Letters}}
  \textbf{\bibinfo{volume}{41}}, \bibinfo{pages}{4328--4331}
  (\bibinfo{year}{2016}).

\bibitem{Juffmann2016a}
\bibinfo{author}{Juffmann, T.}, \bibinfo{author}{Klopfer, B.~B.},
  \bibinfo{author}{Frankort, T. L.~I.}, \bibinfo{author}{Haslinger, P.} \&
  \bibinfo{author}{Kasevich, M.~A.}
\newblock \bibinfo{title}{{Multi-pass microscopy}}.
\newblock \emph{\bibinfo{journal}{Nature Communications}}
  \textbf{\bibinfo{volume}{7}}, \bibinfo{pages}{12858} (\bibinfo{year}{2016}).

\bibitem{Heshmat2018PhotographyDimension}
\bibinfo{author}{Heshmat, B.}, \bibinfo{author}{Tancik, M.},
  \bibinfo{author}{Satat, G.} \& \bibinfo{author}{Raskar, R.}
\newblock \bibinfo{title}{{Photography optics in the time dimension}}.
\newblock \emph{\bibinfo{journal}{Nature Photonics}}
  \textbf{\bibinfo{volume}{12}}, \bibinfo{pages}{560--566}
  (\bibinfo{year}{2018}).

\bibitem{Kollnerr1992}
\bibinfo{author}{K{\"{o}}llner, M.} \& \bibinfo{author}{Wolfrum, J.}
\newblock \bibinfo{title}{{How many photons are necessary for
  fluorescence-lifetime measurements?}}
\newblock \emph{\bibinfo{journal}{Chemical Physics Letters}}
  \textbf{\bibinfo{volume}{200}}, \bibinfo{pages}{199--204}
  (\bibinfo{year}{1992}).

\bibitem{Brinks2015Two-PhotonVoltage}
\bibinfo{author}{Brinks, D.}, \bibinfo{author}{Klein, A.~J.} \&
  \bibinfo{author}{Cohen, A.~E.}
\newblock \bibinfo{title}{{Two-Photon Lifetime Imaging of Voltage Indicating
  Proteins as a Probe of Absolute Membrane Voltage}}.
\newblock \emph{\bibinfo{journal}{Biophysical journal}}
  \textbf{\bibinfo{volume}{109}}, \bibinfo{pages}{914--921}
  (\bibinfo{year}{2015}).

\bibitem{Gong2014ImagingSensors}
\bibinfo{author}{Gong, Y.}, \bibinfo{author}{Wagner, M.~J.},
  \bibinfo{author}{Zhong~Li, J.} \& \bibinfo{author}{Schnitzer, M.~J.}
\newblock \bibinfo{title}{{Imaging neural spiking in brain tissue using
  FRET-opsin protein voltage sensors}}.
\newblock \emph{\bibinfo{journal}{Nature Communications}}
  \textbf{\bibinfo{volume}{5}}, \bibinfo{pages}{3674} (\bibinfo{year}{2014}).

\bibitem{Laviv2016SimultaneousProteins}
\bibinfo{author}{Laviv, T.} \emph{et~al.}
\newblock \bibinfo{title}{{Simultaneous dual-color fluorescence lifetime
  imaging with novel red-shifted fluorescent proteins}}.
\newblock \emph{\bibinfo{journal}{Nature Methods}}
  \textbf{\bibinfo{volume}{13}}, \bibinfo{pages}{989--992}
  (\bibinfo{year}{2016}).

\bibitem{Munro2005TowardImaging}
\bibinfo{author}{Munro, I.} \emph{et~al.}
\newblock \bibinfo{title}{{Toward the clinical application of time-domain
  fluorescence lifetime imaging}}.
\newblock \emph{\bibinfo{journal}{Journal of Biomedical Optics}}
  \textbf{\bibinfo{volume}{10}}, \bibinfo{pages}{051403}
  (\bibinfo{year}{2005}).

\bibitem{Sun2010}
\bibinfo{author}{Sun, Y.} \emph{et~al.}
\newblock \bibinfo{title}{{Fluorescence lifetime imaging microscopy for brain
  tumor image-guided surgery.}}
\newblock \emph{\bibinfo{journal}{Journal of biomedical optics}}
  \textbf{\bibinfo{volume}{15}}, \bibinfo{pages}{056022}
  (\bibinfo{year}{2010}).

\bibitem{Sun2013}
\bibinfo{author}{Sun, Y.} \emph{et~al.}
\newblock \bibinfo{title}{{Endoscopic fluorescence lifetime imaging for in vivo
  intraoperative diagnosis of oral carcinoma.}}
\newblock \emph{\bibinfo{journal}{Microscopy and microanalysis}}
  \textbf{\bibinfo{volume}{19}}, \bibinfo{pages}{791--8}
  (\bibinfo{year}{2013}).

\bibitem{Valm2017}
\bibinfo{author}{Valm, A.~M.} \emph{et~al.}
\newblock \bibinfo{title}{{Applying systems-level spectral imaging and analysis
  to reveal the organelle interactome}}.
\newblock \emph{\bibinfo{journal}{Nature}} \textbf{\bibinfo{volume}{546}},
  \bibinfo{pages}{162--167} (\bibinfo{year}{2017}).

\bibitem{Fan2018Lifetime-engineeredImaging}
\bibinfo{author}{Fan, Y.} \emph{et~al.}
\newblock \bibinfo{title}{{Lifetime-engineered NIR-II nanoparticles unlock
  multiplexed in vivo imaging}}.
\newblock \emph{\bibinfo{journal}{Nature Nanotechnology}}
  \textbf{\bibinfo{volume}{13}}, \bibinfo{pages}{941--946}
  (\bibinfo{year}{2018}).

\bibitem{Niehorster2016a}
\bibinfo{author}{Niehorster, T.} \emph{et~al.}
\newblock \bibinfo{title}{{Multi-target spectrally resolved fluorescence
  lifetime imaging microscopy}}.
\newblock \emph{\bibinfo{journal}{Nat Meth}} \textbf{\bibinfo{volume}{13}},
  \bibinfo{pages}{257--262} (\bibinfo{year}{2016}).

\bibitem{Wang2013LifetimeTrap}
\bibinfo{author}{Wang, Q.} \& \bibinfo{author}{Moerner, W.~E.}
\newblock \bibinfo{title}{{Lifetime and Spectrally Resolved Characterization of
  the Photodynamics of Single Fluorophores in Solution Using the Anti-Brownian
  Electrokinetic Trap}}.
\newblock \emph{\bibinfo{journal}{J. Phys. Chem. B.}}
  \textbf{\bibinfo{volume}{117}}, \bibinfo{pages}{4641--4648}
  (\bibinfo{year}{2013}).

\bibitem{Zhang2015Ultrahigh-throughputMicroscopy}
\bibinfo{author}{Zhang, Z.}, \bibinfo{author}{Kenny, S.~J.},
  \bibinfo{author}{Hauser, M.}, \bibinfo{author}{Li, W.} \&
  \bibinfo{author}{Xu, K.}
\newblock \bibinfo{title}{{Ultrahigh-throughput single-molecule spectroscopy
  and spectrally resolved super-resolution microscopy}}.
\newblock \emph{\bibinfo{journal}{Nature Methods}}  (\bibinfo{year}{2015}).

\bibitem{Mlodzianoski2016Super-ResolutionFluctuations}
\bibinfo{author}{Mlodzianoski, M.~J.}, \bibinfo{author}{Curthoys, N.~M.},
  \bibinfo{author}{Gunewardene, M.~S.}, \bibinfo{author}{Carter, S.} \&
  \bibinfo{author}{Hess, S.~T.}
\newblock \bibinfo{title}{{Super-Resolution Imaging of Molecular Emission
  Spectra and Single Molecule Spectral Fluctuations}}.
\newblock \emph{\bibinfo{journal}{PLoS ONE}} \textbf{\bibinfo{volume}{11}},
  \bibinfo{pages}{0147506} (\bibinfo{year}{2016}).

\bibitem{Squires2017DirectC-phycocyanin}
\bibinfo{author}{Squires, A.~H.} \& \bibinfo{author}{Moerner, W.~E.}
\newblock \bibinfo{title}{{Direct single-molecule measurements of
  phycocyanobilin photophysics in monomeric C-phycocyanin}}.
\newblock \emph{\bibinfo{journal}{PNAS}} \textbf{\bibinfo{volume}{114}},
  \bibinfo{pages}{9779--9784} (\bibinfo{year}{2017}).

\bibitem{Yan2018SpectrallySpectroscopy}
\bibinfo{author}{Yan, R.}, \bibinfo{author}{Moon, S.}, \bibinfo{author}{Kenny,
  S.~J.} \& \bibinfo{author}{Xu, K.}
\newblock \bibinfo{title}{{Spectrally Resolved and Functional Super-resolution
  Microscopy via Ultrahigh-Throughput Single-Molecule Spectroscopy}}.
\newblock \emph{\bibinfo{journal}{Acc. Chem. Res.}}
  \textbf{\bibinfo{volume}{51}}, \bibinfo{pages}{697--705}
  (\bibinfo{year}{2018}).

\bibitem{Dong2016}
\bibinfo{author}{Dong, B.} \emph{et~al.}
\newblock \bibinfo{title}{{Super-resolution spectroscopic microscopy via photon
  localization}}.
\newblock \emph{\bibinfo{journal}{Nature Communications}}
  \textbf{\bibinfo{volume}{7}}, \bibinfo{pages}{12290} (\bibinfo{year}{2016}).

\bibitem{Cordina2018ReducedChelates}
\bibinfo{author}{Cordina, N.~M.} \emph{et~al.}
\newblock \bibinfo{title}{{Reduced background autofluorescence for cell imaging
  using nanodiamonds and lanthanide chelates}}.
\newblock \emph{\bibinfo{journal}{Scientific Reports}}
  \textbf{\bibinfo{volume}{8}}, \bibinfo{pages}{4521} (\bibinfo{year}{2018}).

\bibitem{Fickler2016Quantum10010}
\bibinfo{author}{Fickler, R.}, \bibinfo{author}{Campbell, G.},
  \bibinfo{author}{Buchler, B.}, \bibinfo{author}{Lam, P.~K.} \&
  \bibinfo{author}{Zeilinger, A.}
\newblock \bibinfo{title}{{Quantum entanglement of angular momentum states with
  quantum numbers up to 10,010}}.
\newblock \emph{\bibinfo{journal}{PNAS}} \textbf{\bibinfo{volume}{113}},
  \bibinfo{pages}{13642--13647} (\bibinfo{year}{2016}).

\bibitem{Hahn2012Laser-InducedFields}
\bibinfo{author}{Hahn, D.~W.} \& \bibinfo{author}{Omenetto, N.}
\newblock \bibinfo{title}{{Laser-Induced Breakdown Spectroscopy (LIBS), Part
  II: Review of Instrumental and Methodological Approaches to Material Analysis
  and Applications to Different Fields}}.
\newblock \emph{\bibinfo{journal}{Applied Spectroscopy}}
  \textbf{\bibinfo{volume}{66}}, \bibinfo{pages}{347--419}
  (\bibinfo{year}{2012}).

\bibitem{Gao2016}
\bibinfo{author}{Gao, L.} \& \bibinfo{author}{Wang, L.~V.}
\newblock \bibinfo{title}{{A review of snapshot multidimensional optical
  imaging: measuring photon tags in parallel.}}
\newblock \emph{\bibinfo{journal}{Physics reports}}
  \textbf{\bibinfo{volume}{616}}, \bibinfo{pages}{1--37}
  (\bibinfo{year}{2016}).

\bibitem{Liang2018Single-shotImaging}
\bibinfo{author}{Liang, J.} \& \bibinfo{author}{Wang, L.~V.}
\newblock \bibinfo{title}{{Single-shot ultrafast optical imaging}}.
\newblock \emph{\bibinfo{journal}{Optica}} \textbf{\bibinfo{volume}{5}},
  \bibinfo{pages}{1113--1127} (\bibinfo{year}{2018}).

\bibitem{Jo}
\bibinfo{author}{Jo, S.} \emph{et~al.}
\newblock \bibinfo{title}{{High resolution three-dimensional flash LIDAR system
  using a polarization modulating Pockels cell and a micro-polarizer CCD
  camera}}.
\newblock \emph{\bibinfo{journal}{Optics Express}}
  \textbf{\bibinfo{volume}{24}}, \bibinfo{pages}{1580--1585}
  (\bibinfo{year}{2016}).

\bibitem{Zhang2017HighModulation}
\bibinfo{author}{Zhang, P.} \emph{et~al.}
\newblock \bibinfo{title}{{High resolution flash three-dimensional LIDAR
  systems based on polarization modulation}}.
\newblock \emph{\bibinfo{journal}{Applied Optics}}
  \textbf{\bibinfo{volume}{56}}, \bibinfo{pages}{3889--3894}
  (\bibinfo{year}{2017}).

\bibitem{Chen2018Polarization-modulatedModulator}
\bibinfo{author}{Chen, Z.}, \bibinfo{author}{Liu, B.~O.},
  \bibinfo{author}{Wang, S.} \& \bibinfo{author}{Liu, E.}
\newblock \bibinfo{title}{{Polarization-modulated three-dimensional imaging
  using a large-aperture electro-optic modulator}}.
\newblock \emph{\bibinfo{journal}{Applied Optics}}
  \textbf{\bibinfo{volume}{57}}, \bibinfo{pages}{7750--7757}
  (\bibinfo{year}{2018}).

\end{thebibliography}


\end{document}